\documentclass[aps,twocolumn,PRL]{revtex4}

\usepackage{graphics}
\usepackage{graphicx}
\usepackage{bm}
\usepackage{latexsym}
\usepackage{color, soul}
\usepackage{mathrsfs}
\usepackage{braket}
\usepackage{enumerate}
\usepackage[usenames,dvipsnames]{xcolor}
\usepackage[colorlinks=true,citecolor=blue,linkcolor=blue,urlcolor=blue,backref=true]{hyperref}
\usepackage{float}
\usepackage[raggedright,tight]{subfigure}  
\usepackage{amsmath}
\usepackage{amsfonts}
\usepackage{braket}
\usepackage[normalem]{ulem}
\usepackage{cancel}
\usepackage{mathtools}
\usepackage{appendix}

\begin{document}

\title{Cauchy-Schwarz bound on the accuracy of truncated models in non-relativistic quantum electrodynamics}

\author{Daniel Eyles$^{1,3}$}
\author{Adam Stokes$^2$}
\author{Ahsan Nazir$^1$}

\affiliation{$^1$Department of Physics and Astronomy, University of Manchester, Oxford Road, Manchester M13 9PL, United Kingdom}
\affiliation{$^2$School of Mathematics, Statistics, and Physics, Newcastle University, Newcastle upon Tyne NE1 7RU, United Kingdom}
\affiliation{$^3$Institute of Materials Research and Engineering, A*STAR (Agency for Science, Technology and Research), 2 Fusionopolis Way, Innovis \#08-03, 138634 Singapore, Republic of Singapore}

\begin{abstract}
We show that the Cauchy-Schwarz inequality provides a simple yet general bound that limits the accuracy of light-matter theories which retain only finite numbers of material energy levels. A corollary is that unitary rotations within a truncated space cannot transform between gauges, because the contrary assumption yields incorrect predictions. In particular, a widespread model obtained using such a rotation and treated as a Coulomb gauge model within a substantial body of literature, yields incorrect predictions under this assumption. The simplest system of a single dipole coupled to a single photonic mode in one spatial dimension is analysed in detail.
\end{abstract}

\maketitle

{\em Introduction}. Models that include only two material energy levels, such as the quantum Rabi model (QRM) \cite{PhysRev.49.324} and the Jaynes Cummings model \cite{JC_Model_Origional}, have been instrumental in shaping our understanding of quantum light-matter physics. But because in quantum electrodynamics the canonical quantum subsystems conventionally referred to as ``matter" and ``photons" are physically distinct in different gauges, material truncation breaks gauge-invariance \cite{de_bernardis_breakdown_2018,stokes_gauge_2019}. Truncated theories can nevertheless remain accurate even beyond conventional weak-coupling regimes, and a number of them have now been studied in various settings \cite{roth_optimal_2019,de_bernardis_breakdown_2018,di_stefano_resolution_2019,gustin_gauge-invariant_2023,ashida_cavity_2021,stokes_implications_2022,taylor_resolution_2020,stokes_gauge_2019,li_electromagnetic_2020}.

The optimal gauge for truncation generally depends on the system and regime considered and the physical prediction sought, and it does not necessarily coincide with the gauge relative to which the most operationally relevant definitions of matter and photons are obtained \cite{stokes_implications_2022}. Numerous properties of interest have been studied under material truncation, including global energy spectra \cite{roth_optimal_2019,de_bernardis_breakdown_2018,di_stefano_resolution_2019,li_electromagnetic_2020,ashida_cavity_2021,stokes_implications_2022,taylor_resolution_2020,stokes_gauge_2019}, emission spectra \cite{salmon_gauge-independent_2022}, as well as subsystem properties such as photodetection amplitudes \cite{settineri_gauge_2021,gustin_gauge-invariant_2023}, and light-matter entanglement \cite{arwas_metrics_2023,stokes_gauge-relativity_2023}. However, until now a general quantitative bound on the accuracy of truncated theories has not been given.

Our results reveal the relative significance of different truncated theories and the methods used to derive them, which has been a subject of debate. A gauge-principle can be defined within a truncated space using the corresponding truncated position operator \cite{di_stefano_resolution_2019,savasta_gauge_2021}, and despite arguments to the contrary \cite{stokes_gauge_2020,stokes_gauge_2024,stokes_implications_2022}, this has been assumed to yield accurate truncated theories in gauges that otherwise seem not to possess them, such as the Coulomb gauge. In particular, a now widespread Hamiltonian $h_1(0)$ derived in Ref.~\cite{di_stefano_resolution_2019} and equivalent to a dipole gauge truncated Hamiltonian, was supposed therein to offer an accurate Coulomb gauge model. This supposition has subsequently been carried forward into what is now a substantial body of literature \cite{akbari_generalized_2023, dmytruk_gauge_2021, gustin_gauge-invariant_2023, gustin_what_2024,hughes_reconciling_2024, mercurio_flying_2022, mercurio_pure_2023, mercurio_regimes_2022, nodar_identifying_2023, salmon_gauge-independent_2022, savasta_gauge_2021, settineri_gauge_2021,garziano_gauge_2020,taylor_resolution_2020,mandal_theoretical_2023,taylor_resolving_2022,le_boite_theoretical_2020}, and it has played a key role in the calculation and interpretation of predictions therein. Yet the supposition constitutes a tacit equating of two separate gauge freedoms \cite{stokes_implications_2022}. Whether it is actually valid is addressed in this letter, with negative outcome.

We use the Cauchy-Schwarz inequality (CSI) to formalise the intuitive idea that a normalised vector within a Hilbert space ${\cal H}$ cannot be replicated by a normalised vector within a subspace $P{\cal H}$ defined by a projection $P$, if it has a non-negligible component within the orthogonal complement $Q{\cal H}$ where $Q=I-P$. An implication is that it is impossible in certain gauges, such as the Coulomb gauge, to replicate the vectors and operators that represent physical states and observables of interest, using any theory that includes only a few low-lying material energy levels. As an immediate corollary, we show that frames within a truncated space connected by unitary rotations cannot be identified with gauges of the non-truncated theory, otherwise, incorrect predictions are obtained. In particular, the $i$'th eigenvector $\ket{e_1^i(0)}$ of $h_1(0)$ cannot replicate the corresponding exact eigenvector $\ket{E_0^i}$ of the Coulomb gauge Hamiltonian $H_0$, such that $h_1(0)$ cannot approximate $H_0$. More generally, the average $\langle {\cal O}\rangle_i$ of some observable ${\cal O}$ in an energy eigenstate ${\cal S}_i$ is given exactly by $\bra{E_0^i}O_0\ket{E_0^i}$, from which $\bra{e_1^i(0)}O_0\ket{e_1^i(0)}$ generally deviates significantly.

{\em Material truncation}. The Hilbert space and operator algebra of non-relativistic quantum electrodynamics can be partitioned as ${\cal H} = {\cal H}_m\otimes {\cal H}_{\rm ph}$ and  ${\cal A} = {\cal A}_m\otimes {\cal A}_{\rm ph}$. The pairs $({\cal H}_m, {\cal A}_m)$ and $({\cal H}_{\rm ph},{\cal A}_{\rm ph})$ define canonical quantum subsystems called matter and photons respectively. The vector and operator representations of state $\cal S$ and observable ${\cal O}$ in gauges $g$ and $g'$ are connected by a unitary gauge-fixing transformation $U_{gg'}$ as $\ket{S_{g'}}=U_{gg'}\ket{S_g}$ and $O_{g'}=U_{gg'}O_g U_{gg'}^\dagger$ respectively \cite{stokes_implications_2022}. Physical predictions are unique (gauge-invariant), $\bra{S_g}O_g\ket{S_g}=\langle {\cal O}\rangle_{\cal S} = \bra{S_{g'}}O_{g'}\ket{S_{g'}}$, but since $U_{gg'}\neq U_m\otimes U_{\rm ph}$, in each different gauge the matter and photon subsystems are physically distinct \cite{stokes_implications_2022}.

The Hamiltonian operator represents the total observable energy $E$ and can be partitioned in an arbitrary gauge $g$ as $H_g=H_m \otimes I_{\rm ph} +I_m\otimes H_{\rm ph}+V_g$ where $H_m \in {\cal A}_m$ and $H_{\rm ph}\in {\cal A}_{\rm ph}$ are material and photonic bare energies respectively. We denote the eigenvalues and eigenvectors of $H_m$ by $\epsilon_\mu$ and $\ket{\epsilon_\mu}$ respectively. It is common in order to obtain simpler theories to replace ${\cal H}$ with a truncation $P{\cal H}$ where $P=\sum_{\mu=0}^M \ket{\epsilon_\mu}\bra{\epsilon_\mu}\otimes I_{\rm ph}$ with $M<\infty$. However, doing so results in non-equivalent theories in different gauges \cite{stokes_implications_2022}.\vspace*{1mm}

{\em Cauchy-Schwarz bound}\label{cssec}.
\begin{figure}
\begin{minipage}{\columnwidth}
\begin{center}
\vspace*{1mm}
\hspace*{0.5mm}\includegraphics[scale=0.20]{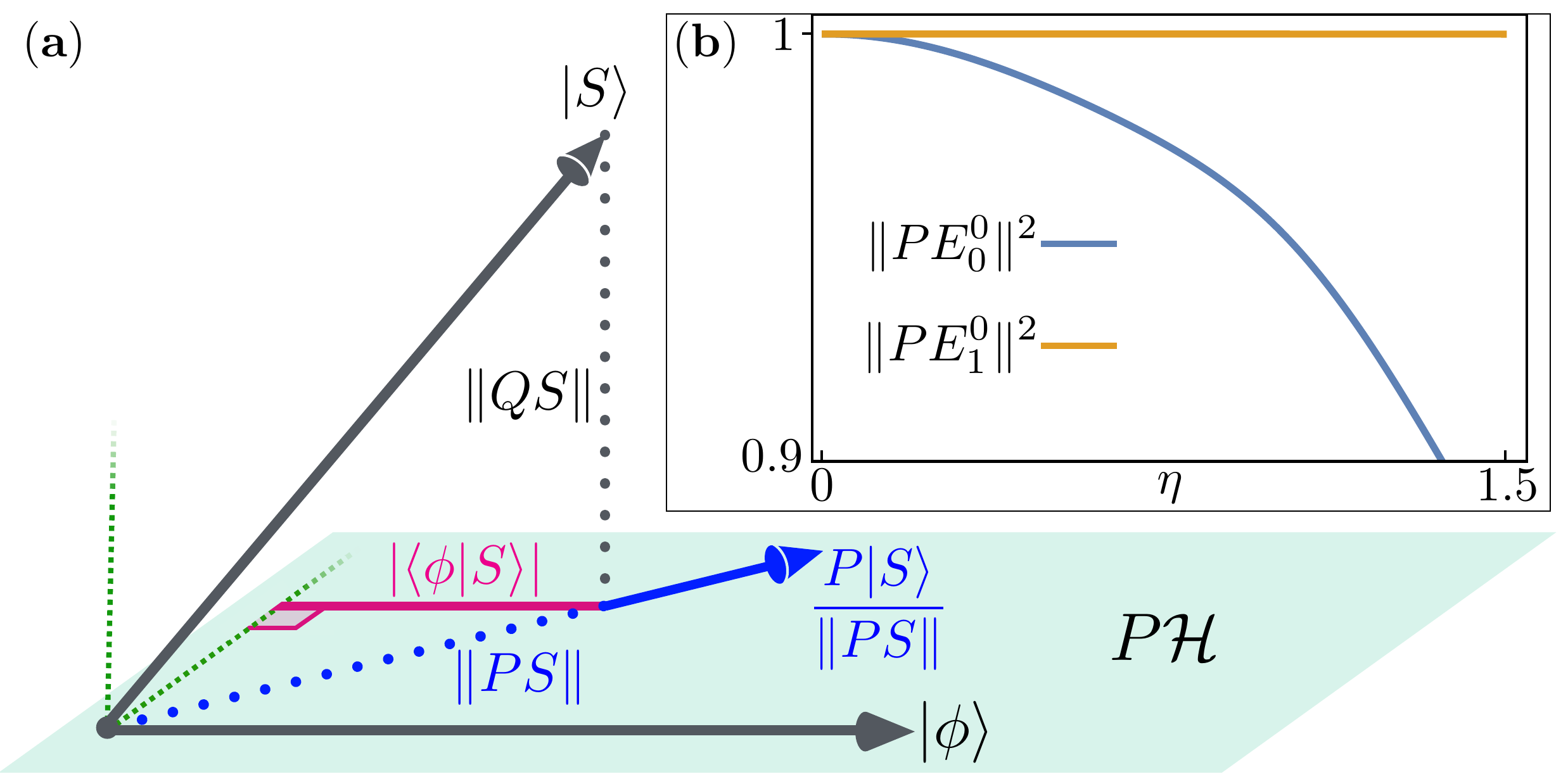}
\caption{\textbf{(a)} The normalised vector $\ket{\phi}$ lying in the plane $P{\cal H}$ (shaded) is shown with the normalised vector $\ket{S}$. Green dashed lines show directions orthogonal to $\ket{\phi}$. The CSI implies that the length of the component of $\ket{S}$ along $\ket{\phi}$, namely $|\langle \phi| S\rangle |$ (solid magenta), is less than or equal to the length $\Vert PS \Vert$ (dotted blue) of the component of $\ket{S}$ in the plane $P{\cal H}$. In short, $|\langle \phi | S\rangle|= |\langle \phi |P| S\rangle|\leq \Vert \phi \Vert \Vert PS \Vert = \Vert PS \Vert $. This bound is saturated if $\ket{\phi}=P\ket{ S}/\Vert PS \Vert$ (blue, dotted $+$ solid). \textbf{(b)} For a single-mode anharmonic dipole system under two-level material truncation, ground-state fidelity upper bounds  $\|PE_0^0\|^2$ (Coulomb gauge) and $\|PE_1^0\|^2$ (dipole gauge), are plotted with coupling strength $\eta$.}\label{gpic}
\end{center}
\end{minipage}
\end{figure}
The length of a vector in ${\cal H}$ is defined using the inner-product as $\Vert{\cdot}\Vert^2:=\braket{\cdot |\cdot}$. We will now show that the CSI $|\braket{u|v}|\leq \Vert{u}\Vert\Vert{v}\Vert~~\forall\,\ket{u},\,\ket{v}\in {\cal H}$,  places an upper bound on how well a truncated theory can replicate exact state vectors. We define the fidelity $F(u, v) := |\braket{u|v}|^2$, let projection $P:{\cal H}\to P{\cal H}$ define a subspace $P{\cal H} \subset {\cal H}$, and let $\ket{S}\in {\cal H}$ and $\ket{\phi} \in P{\cal H}$ be normalised. Since $F(\phi, S)\equiv  |\braket{\phi|P|S}|^2$, the CSI implies
\begin{align}\label{bound3}
F(\phi,S) \leq \Vert PS \Vert^2 = 1- \Vert QS \Vert^2,
\end{align}
where $Q=I-P$.~In words, whenever $\ket{S}$ possesses a non-negligible component within $Q{\cal H}$ the fidelity $F(\phi, S)$ will exhibit non-negligible deviations from unity for {\em any} normalised $\ket{\phi} \in P{\cal H}$, as illustrated in Fig.~\ref{gpic}(a). Conversely, $F(\phi, S)\approx 1$ implies that $\ket{z} :=\ket{S}-\braket{\phi|S}\ket{\phi}$ is such that $\braket{z|z}=1-F(\phi,S)\approx 0$, from which it follows by definition of the inner-product that $\ket{z}$ is the zero vector, such that $\ket{S} \approx \braket{\phi|S}\ket{\phi}$. Since $F(\phi, S)\approx 1$ implies $|\braket{\phi|S}|\approx 1$ it follows that $\ket{S} \approx \ket{\phi}$ up to an ignorable phase. We can now cast our bound in terms of eigenvalues. Suppose that $O\in {\cal A}$ satisfies $O \ket{O^i}= O^i\ket{O^i}$ with $\Vert O^i \Vert =1$. If there exists $\ket{\phi}\in P{\cal H}$ such that $F(\phi, O^i)\approx 1$ then by bound (\ref{bound3}) we have $1\approx \|PO^i\|^2 = F(PO^i/\|PO^i\|,O^i)$. It follows that $\|PO^i\|\approx 1$ and that up to a phase $\ket{O^i}\approx P\ket{O^i}/\|PO^i\| \approx P\ket{O^i}$. Noting that $P^2=P$ it follows in turn that $PO P P \ket{O^i} \approx O^i P\ket{O^i}$.~Thus, if the operator $PO P$ possesses no eigenvalues that approximate $O^i$, then there exists no normalised vector $\ket{\phi}\in P{\cal H}$ that approximates $\ket{O^i}$. 

{\em Gauges versus frames within $P{\cal H}$}. A truncating map on ${\cal A}$ is any map $M_P: {\cal A}\to {\cal A}_P$ where ${\cal A}_P$ denotes the algebra of Hermitian operators over $P{\cal H}$. Suppose that a gauge $g$ admits an accurate truncation in the sense that each state ${\cal S}$ of interest represented by a vector $\ket{S}\in{\cal H}$, also admits a representation $\ket{S_{g,P}}\in P{\cal H}$, because $F(S_{g,P},S)\approx 1$. In this gauge $M_P(O_g)=PO_gP$ is a permissible truncation of $O_g$ representing ${\cal O}$, because
\begin{align}
\langle {\cal O}\rangle_{\cal S} &= \bra{S_g}O_g\ket{S_g} \nonumber \\ &\approx  \bra{S_{g,P}}O_g\ket{S_{g,P}} = \bra{S_{g,P}}PO_gP\ket{S_{g,P}}.\label{approxav}
\end{align}
Of particular interest within the literature have been averages in the $i'$th energy eigenstate ${\cal S}_i$ represented by eigenvector $\ket{E_g^i}$ of $H_g$. Eq.~(\ref{approxav}) then reads $\langle {\cal O}\rangle_i  \approx \bra{E^i_{g,P}}PO_gP\ket{E^i_{g,P}}$ where $\ket{E_{g,P}^i}\approx \ket{E_g^i}$ is the $i$'th eigenvector of an accurate truncation $M_P(H_g) \in {\cal A}_P$. 

Once a gauge $g$ admitting an accurate truncation has been identified, {\em any} unitary ${\cal U}:P{\cal H}\to P{\cal H}$ can trivially be used to define equivalent truncated representations of ${\cal S}$ and ${\cal O}$ as ${\cal U}\ket{S_{g,P}}$ and ${\cal U}PO_gP{\cal U}^\dagger$ respectively.~This construction relies entirely upon the gauge $g$ from which the new representations must be generated, and so it merely offers a more circuitous route to obtaining the same predictions already given directly by Eq.~(\ref{approxav}). A unitary ${\cal T}_{gg'}:P{\cal H}\to P{\cal H}$ could be considered {\em useful} only if it were to generate an accurate truncated theory within a different gauge $g'$ by allowing $\langle {\cal O}\rangle_{\cal S}$ to be calculated as 
 \begin{align}
\langle {\cal O}\rangle_{\cal S} \approx  \bra{s_{g',P}}O_{g'}\ket{s_{g',P}} = \bra{s_{g',P}}PO_{g'}P\ket{s_{g',P}} \label{approxav2}
\end{align} 
where
$\ket{s_{g',P}}={\cal T}_{gg'}\ket{S_{g,P}}$ and $O_{g'}$ represents ${\cal O}$ in the gauge $g'$. For example, once $h_g(g') = {\cal T}_{gg'}M_P(H_g){\cal T}_{gg'}^\dagger$ were obtained, 
{\em any} average of the form $\langle {\cal O}\rangle_i$ could be calculated autonomously, that is, without further reliance upon gauge $g$, by using Eq.~(\ref{approxav2}) and letting $\ket{s_{g',P}}$ be the $i$'th eigenvector $\ket{e^i_g(g')}$ of $h_g(g')$.

However, inequality (\ref{bound3}) implies that if $\ket{S_{g'}}=U_{gg'}\ket{S_g}$ possesses a non-negligible component within $Q{\cal H}$ then it cannot be accurately represented by any vector in $P{\cal H}$. Noting that $\langle {\cal O}\rangle_{\cal S} =  \bra{S_{g'}}O_{g'}\ket{S_{g'}}$, the approximation in Eq.~(\ref{approxav2}) is seen to be invalid, and there exists no transformation ${\cal T}_{gg'}:P{\cal H}\to P{\cal H}$ that can give a generally accurate truncated theory {\em in the gauge} $g'$. The contrary supposition is predominant within the literature \cite{akbari_generalized_2023, dmytruk_gauge_2021, gustin_gauge-invariant_2023, gustin_what_2024,hughes_reconciling_2024, mercurio_flying_2022, mercurio_pure_2023, mercurio_regimes_2022, nodar_identifying_2023, salmon_gauge-independent_2022, savasta_gauge_2021, settineri_gauge_2021,garziano_gauge_2020,taylor_resolution_2020,mandal_theoretical_2023,taylor_resolving_2022,le_boite_theoretical_2020}. The construction of models $h_g(g')$ using certain ${\cal T}_{gg'}:P{\cal H}\to P{\cal H}$ has been proposed as a fundamental gauge-principle \footnote{The gauge-principle within $P{\cal H}$ defined using the operator $PxP$ is satisfied by every class ${\cal C}_\alpha$. Such a principle may find use in describing {\em ab initio} finite systems like lattices, but its automatic satisfaction by ${\cal C}_\alpha$ does not guarantee accurate predictions of the models therein, and it does not distinguish any one class ${\cal C}_\alpha$ from the others. A distinguishing feature of the class ${\cal C}_1$ is that it is derivable via unitary rotations of the free Hamiltonians $H_m$ and $H_{\rm ph}$, unlike other ${\cal C}_\alpha$. This is a separate property, rather than a requirement of any known gauge principle. Indeed, the property cannot be satisfied in QED without approximations, being admitted by ${\cal C}_1$ only within the electric dipole approximation \cite{stokes_gauge_2024}.} within $P{\cal H}$ analogous to that encountered in lattice theories \cite{di_stefano_resolution_2019,savasta_gauge_2021}. In particular, a model $h_1(0)$ introduced in Ref.~\cite{di_stefano_resolution_2019} (defined by $g=1$ and $g'=0$) has been taken to be a Coulomb gauge ($g=0$) model. 
The failure of this assumption is demonstrated with several examples in what follows.
\vspace*{1mm}

{\em Toy model: A single dipole and single mode}. We consider a charge $q$ with position $x\in {\cal A}_m$ bound in a potential $V(x)$, interacting with a single mode described by the one-dimensional amplitude $A$ of the gauge-invariant transverse vector potential, such that $-{\dot A}=E_{\rm T}$ is the corresponding amplitude of the transverse electric field. Canonical momenta $p \in {\cal A}_m$ and $\Pi \in {\cal A}_{\rm ph}$ satisfy $[x,p] = i$ and $[A,\Pi] = i/v$. Photonic mode expansions of $A$ and $\Pi$ read $A={1\over \sqrt{2\omega v}}(a+a^\dagger)$ and $\Pi=i\sqrt{\omega\over 2v}(a^\dagger -a)$ with $[a,a^\dagger]=1$. Within ${\cal A}$, any material (photonic) operator includes a tensor product with $I_{\rm ph}$ ($I_m$), which we henceforth understand as implicit.

The choice of gauge is encoded into a single real parameter $\alpha$ and the Hamiltonian in any gauge is the sum of mechanical and transverse electromagnetic Hamiltonians; \cite{stokes_implications_2022,stokes_gauge_2024,stokes_gauge_2020} $H_\alpha = \mathscr{H}_{\rm mech,\alpha} + \mathscr{H}_{\rm TEM,\alpha}$, where $\mathscr{H}_{\rm mech,\alpha} = {1\over 2}m{\dot x}^2+V(x)$ and $\mathscr{H}_{\rm TEM,\alpha} = {v\over 2}\big[E_{\rm T}^2 + \omega^2 A^2 \big]$ in which $m{\dot x} = -im[x,H_\alpha]=p -q(1-\alpha)A$ and $-E_{\rm T} = {\dot A} = -i[A,H_\alpha]=\Pi+\alpha x/v$. The choices $\alpha=0$ and $\alpha=1$ are called the Coulomb gauge and dipole gauge respectively.~The unitary gauge-fixing transformation between gauges $\alpha$ and $\alpha'$ is $R_{\alpha\alpha'} = \exp\left[ iq(\alpha - \alpha')xA \right]$, such that $X_{\alpha'}=R_{\alpha\alpha'}X_\alpha R_{\alpha\alpha'}^\dagger$ with $X=H,\,\mathscr{H}_{\rm mech},\,\mathscr{H}_{\rm TEM}$ \cite{stokes_implications_2022,stokes_gauge_2024,stokes_gauge_2020}. 

The bare matter and photon Hamiltonians are $H_m=p^2/(2m)+V(x)$ and $H_{\rm ph}=\omega(a^\dagger a +1/2)$ respectively. Hereafter we consider the generic example of a double-well dipole with potential  $V(x) = -\theta x^2/2 + \phi x^4/4$ where $\phi$ and $\theta$ fix the shape of the double well. 
We consider the case of resonance $\omega/\omega_0 =  1$, where $\omega_0:=\epsilon_1-\epsilon_0$, and choose $\theta$ and $\phi$ to give a very high material anharmonicity of $\mu \approx 70$ where $\mu = (\omega_{1} - \omega_0)/\omega_0$ in which $\omega_{1}=\epsilon_2-\epsilon_1$. To best make contact with existing literature we will consider a two-level material truncation ($M=1$) for which $P=\ket{\epsilon_0}\bra{\epsilon_0}+\ket{\epsilon_1}\bra{\epsilon_1}$. It is well-known that in this case the dipole gauge ($\alpha=1$) and Coulomb gauge ($\alpha=0$) produce relatively accurate and inaccurate standard truncations respectively \cite{de_bernardis_breakdown_2018,stokes_implications_2022,di_stefano_resolution_2019,roth_optimal_2019}.

We illustrate inequality (\ref{bound3}) by considering energy eigenstates ${\cal S}_i$. In the gauge $\alpha$ the energy $E$ is represented by $H_\alpha$ whose $i$'th eigenvector we denote by $\ket{E_\alpha^i}$. The upper bound (\ref{bound3}) is $\Vert PE_\alpha^i \Vert^2 = F(\epsilon_0,E_\alpha^i)+F(\epsilon_1,E_\alpha^i)$. Figure \ref{gpic} shows the ground state upper bound $\Vert PE_\alpha^0 \Vert^2$ for the Coulomb ($\alpha=0$) and dipole ($\alpha=1$) gauges as a function of the dimensionless coupling strength $\eta  := qx_{10}/\sqrt{2\omega v}$ in which $x_{10} := \bra{\epsilon_1}x\ket{\epsilon_0}$ is assumed to be real.  The eigenvector $\ket{E_1^0}$ does essentially reside within $P{\cal H}$ for all coupling strengths shown, but this is not true of $\ket{E_0^0}$. Inequality (\ref{bound3}) implies that this behaviour could also have been deduced by noting that the Coulomb gauge truncated model $H_0^2:= PH_0P$ is known to produce an inaccurate energy spectrum for sufficiently large $\eta$ \cite{de_bernardis_breakdown_2018,stokes_implications_2022,di_stefano_resolution_2019,stokes_gauge_2019}. In Supplementary Material (SM) A we compare the ground state fidelities achieved by noteworthy dipole gauge two-level models with the corresponding upper bound.


{\em Predictions of $h_1(0)$ treated as a Coulomb gauge model}. A commonly used truncating map $M_P$ which we call the ``standard" truncating map is such that there exists a gauge, namely $\alpha=1$, for which $M_P(H_1)$ is the paradigmatic standard QRM. It is defined by $
M_P(H_\alpha) = PH_mP + PH_{ph}P + V_\alpha(PxP,PpP) =: H_\alpha^2$ where $V_\alpha:=H_\alpha-H_m-H_{\rm ph}$. Note that $H_\alpha^2=PH_\alpha P$ if and only if $\alpha=0$. The models $H_\alpha^2$ are not equivalent for different $\alpha$. For each $\alpha$, an equivalence class can be generated from $H_\alpha^2$ as ${\cal C}_\alpha:=\{h_\alpha(\alpha'):\alpha'\in {\mathbb R}\}$, where $h_\alpha(\alpha') := \mathcal{T}_{\alpha\alpha'} H_\alpha^2 \mathcal{T}_{\alpha\alpha'}^\dagger$ in which $\mathcal{T}_{\alpha\alpha'} :=\exp\left[ iq(\alpha - \alpha')PxPA \right]$ is a truncated space analog of $R_{\alpha\alpha'}$. The apparent lack of a truncated model accurate beyond the weak-coupling regime within gauges close to the Coulomb gauge ($\alpha=0$) has been perceived as a problem that requires ``resolution" \cite{di_stefano_resolution_2019,taylor_resolution_2020}. In the context of a single anharmonic dipole and single mode, the model $h_1(0)\in {\cal C}_1$ has now received relatively widespread attention \cite{akbari_generalized_2023, dmytruk_gauge_2021, gustin_gauge-invariant_2023, gustin_what_2024,hughes_reconciling_2024, mercurio_flying_2022, mercurio_pure_2023, mercurio_regimes_2022, nodar_identifying_2023, salmon_gauge-independent_2022, savasta_gauge_2021, settineri_gauge_2021,garziano_gauge_2020,taylor_resolution_2020,mandal_theoretical_2023,taylor_resolving_2022,le_boite_theoretical_2020}, the proposal being that, {\em i}) it is accurate, because it is equivalent to $H_1^2$, {\em and}, {\em ii}) it is a Coulomb gauge model, because it is the $\alpha=0$ member of the class ${\cal C}_1$. 

We now illustrate the important implication of inequality (\ref{bound3}), that properties {\em i}) and {\em ii}) cannot be simultaneously satisfied. If $h_1(0)$ is treated as a Coulomb gauge model, then in general it yields incorrect predictions. We let $\ket{E_1^{2,i}}$ and $\ket{e_1^i(0)} = {\cal T}_{10}\ket{E_1^{2,i}}$ denote the $i$'th eigenvectors of $H_1^2$ and $h_1(0)$ respectively. Consider observable ${\cal O}$ represented in the gauge $\alpha$ by operator $O_\alpha = R_{\alpha'\alpha}O_{\alpha'}R_{\alpha'\alpha}^\dagger$. The exact average of ${\cal O}$ in the $i$'th energy eigenstate ${\cal S}_i$ is computed in the Coulomb gauge as $\langle {\cal O}\rangle_i = \bra{E_0^i} O_0 \ket{E_0^i}$. It is clear that $h_1(0)$ can only be viewed as a correct Coulomb gauge model if it provides an accurate approximation of this average in the form $\langle {\cal O} \rangle_i \approx \bra{e_1^i(0)} O_0 \ket{e_1^i(0)}$ [cf.~Eq.~(\ref{approxav2})]. It is equally clear however, that this approximation cannot generally hold, because by bound (\ref{bound3}) the eigenstates $\ket{E_0^i}$ cannot be replicated for sufficiently large $\eta$ by {\em any} two-level model (Fig.~\ref{gpic}).

Let us instead consider the average $\langle {\cal O} \rangle_i$ found using the dipole gauge QRM $H_1^2$, which is $\langle {\cal O} \rangle_i \approx \bra{E_1^{2,i}}PO_1 P\ket{E_1^{2,i}}$ [cf.~Eq.~(\ref{approxav})]. Since $O_1 = R_{01}O_0R_{01}^\dagger$ and $R_{01}^\dagger P\ket{E_1^{2,i}} = R_{01}^\dagger \ket{E_1^{2,i}} \approx R_{01}^\dagger \ket{E_1^i} = \ket{E_0^i}$, this average does indeed approximate the exact average. The models $H_1^2$ and $h_1(0)$ are equivalent and the former yields an accurate average whereas the latter fails to do so when identified as a Coulomb gauge model. It is therefore this identification that fails. The correct representation of ${\cal O}$ 
to use in conjunction with $h_1(0)$ is not the Coulomb gauge representation $O_0$. It must instead be constructed from the truncated dipole gauge representation $PO_1P$ as ${\cal T}_{10}PO_1P {\cal T}_{10}^\dagger$. Important examples are now given. 

\begin{figure}[t]
\begin{minipage}{\columnwidth}
\begin{center}
\hspace*{-1mm}\includegraphics[width=1.02\textwidth]{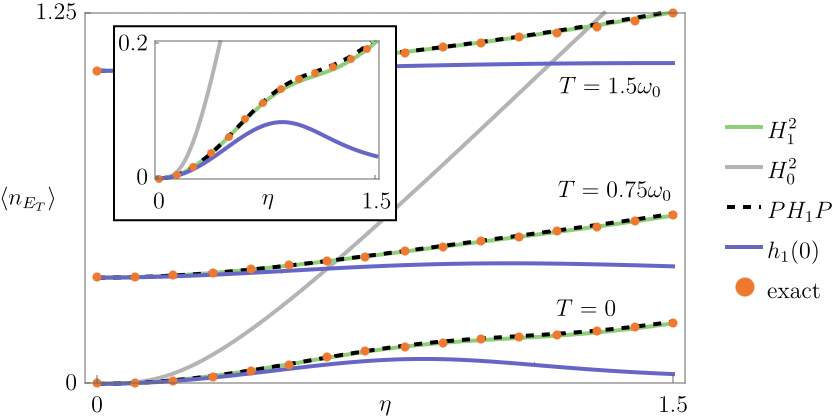}
\caption{Thermal averages $\langle n_{E_{\rm T}}\rangle$ at different temperatures $T$ found using different two-level models and the exact (non-truncated) theory, are plotted with coupling strength $\eta$, with $\omega=\omega_0$ and $\mu\approx 70$. The inset shows the $T=0$ (pure ground state) case wherein the difference between the blue and green solid curves is the average $\bra{E_1^{2,0}}\Delta \ket{E_1^{2,0}}$. 
}\label{ph num}
\vspace*{-2mm}
\end{center}
\end{minipage}
\end{figure}

We first consider the number of $E_{\rm T}$-type photons defined as
\begin{align}\label{phot}
n_{E_{\rm T}} = {v \over 2\omega}(E_{\rm T}^2 + \omega^2 A^2)-{1\over 2}.
\end{align}
This definition of photon has often been tacitly assumed to be of primary relevance in the context of photodetection theory within the ultrastrong coupling regime, and has often been used in conjunction with the eigenstates of $h_1(0)$ \cite{settineri_gauge_2021,gustin_gauge-invariant_2023}. In the Coulomb gauge, $n_{E_{\rm T}}$ is represented by the operator $a^\dagger a$. The average $\langle n_{E_{\rm T}}\rangle_i$ found using $h_1(0)$ treated as a Coulomb gauge model is $\bra{e^i_1(0)}a^\dagger a \ket{e_1^i(0)}$. The difference between this average and the corresponding accurate average found using $H_1^2$ is $\bra{E_1^{2,i}} \Delta \ket{E_1^{2,i}}$ where $\Delta:=\mathcal{T}_{01}a^\dagger a \mathcal{T}_{01}^\dagger - PR_{01}a^\dagger aR_{01}^\dagger P$ is straightforwardly found to be given by
\begin{equation}
    \Delta  = \eta^2 \bigg( {Px^2 P \over (Px P)^2} - I \bigg), \label{diffnet}
\end{equation}
in which $(Px P)^{-2} := x_{10}^{-2}I$. This difference is significant for sufficiently large coupling strengths as shown in Fig.~\ref{ph num}.

One can also consider the total excited state population of the dipole defined relative to the Coulomb gauge, that is, the total population $\Gamma$ of the excited states of $K = {1\over 2m}(m{\dot x}+qA)^2+V(x)$, which is represented in the Coulomb gauge by the operator $H_m$. Results for $\Gamma$ are given in SM B, and mirror those obtained for $n_{E_{\rm T}}$. 
The correct representations of $n_{E_{\rm T}}$ and $K$ within the frame defining $h_1(0)$ are respectively obtained by letting $O_0=a^\dagger a$ and $O_0=H_m$ within ${\cal T}_{10}PR_{01}O_0 R_{01}^\dagger{\cal T}_{10}^\dagger$. We conclude that the frame defining $h_1(0)$ provides altogether different definitions to the Coulomb gauge of both photons and matter.

\begin{figure}[t]
\begin{minipage}{\columnwidth}
\begin{center}
\hspace*{-1mm}\includegraphics[width=0.98\textwidth]{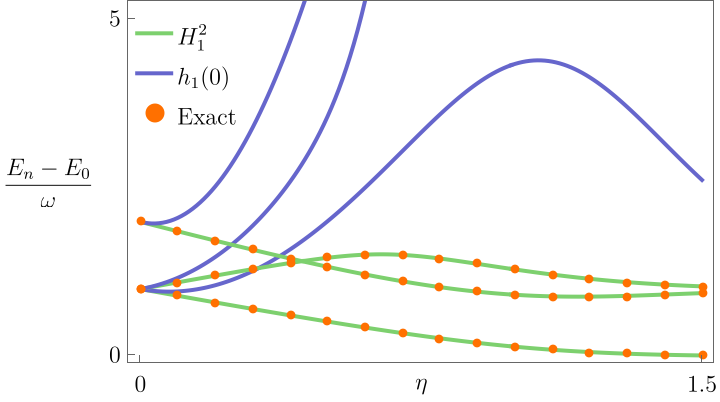}
\caption{The first three transition energies relative to the ground state are compared for the exact theory and the QRM $H_1^2$, with those computed using $h_1(0)$ treated as a Coulomb gauge model as $\bra{e_1^i(0)} H_0 \ket{e_1^i(0)}- \bra{e_1^0(0)} H_0 \ket{e_1^0(0)}$. The parameters chosen are as in Fig. \ref{ph num}}\label{Assume Correct Model}
\vspace*{-2mm}
\end{center}
\end{minipage}
\end{figure}

Finally, we consider a global property, the total energy $E$. Its average in the $i$'th energy eigenstate is the $i$'th energy eigenvalue of $H_\alpha$; $\langle E \rangle_i := \bra{E_\alpha^i}H_\alpha \ket{E_\alpha^i} = E_i$. As is well-known, within the regimes we have considered transition energies relative to the ground energy can be computed accurately using the eigenvalues of $H_1^2$ as $E_i - E_0 \approx \bra{E_1^{2,i}} H_1^2 \ket{E_1^{2,i}}- \bra{E_1^{2,0}}H_1^2 \ket{E_1^{2,0}} = E_1^{2,i}-E_1^{2,0}$ \cite{de_bernardis_breakdown_2018,stokes_implications_2022,di_stefano_resolution_2019,roth_optimal_2019}, which shows that for this purpose $H_1^2$ is both accurate {\em and} that it is a dipole-gauge model. All models $h_1(\alpha)\in{\cal C}_1$ are unitarily equivalent to $H_1^2=h_1(1)$ and therefore trivially possess identical eigenvalues, but since $E_i-E_0 \not\approx \bra{e_1^i(0)} H_0 \ket{e_1^i(0)}- \bra{e_1^0(0)} H_0 \ket{e_1^0(0)}$ (Fig.~\ref{Assume Correct Model}), one sees that $h_1(0)$ does not approximate $H_0$. Indeed, an operator $O$ approximates $H_0$ if and only if the eigenvalues and eigenvectors of $O$ approximate those of $H_0$, but by inequality (\ref{bound3}) the eigenvectors of $H_0$ cannot be reproduced by any two-level model.

In fact, as shown in SM C, the eigenvalues of $h_1(\alpha)\in {\cal C}_1$ also deviate significantly from the exact eigenvalues $E_i$ of $H_\alpha$. In contrast, the eigenvalues $PH_1P$ do accurately approximate the $E_i$, otherwise by bound (\ref{bound3}) the dipole gauge could not admit an accurate truncation. The eigenvalues of models in ${\cal C}_1$ can be used to obtain accurate transition energies (Fig.~\ref{Assume Correct Model}) only because the difference $PH_1P-H_1^2=\omega \Delta$ with $\Delta$ given by Eq.~(\ref{diffnet}), is such that $ \langle \Delta \rangle_i$ is essentially independent of $i$ (see SM C). 

{\em Open dipole-mode system.} We can extend the results above to the case that the dipole mode system is open. A commonly found description of loss characterised by rate $\kappa$ is obtained by coupling a dimensionless observable ${\cal O}$ of the system to an external reservoir with flat spectrum \cite{salmon_gauge-independent_2022}. We let the operator representations of the state ${\cal S}$ at time $t$, the observable ${\cal O}$, and the energy $E$ be denoted $\rho$, $O$ and $H$ respectively. The Lindblad master equation ${\dot \rho} =-i[\rho,H]+{\cal D}(\rho)$ for the density operator $\rho$ possesses dissipator 
\begin{align}\label{me}
&{\cal D}(\rho):=\kappa\sum_{\cal E} \left[O^-({\cal E})\rho O^+({\cal E}) -{1\over 2}\{O^+({\cal E})O^-({\cal E}),\rho\}\right]\nonumber \\
&O^+({\cal E}) = \sum_{\substack{i,j,~i>j \\ E_i-E_j={\cal E}}} \bra{i}O\ket{j}\ket{i}\bra{j},~~~ O^-=(O^+)^\dagger
\end{align}
where $H\ket{i}= E_i\ket{i}$ defines the energy eigenvalues $E_i$, and the vector representations $\ket{i}$ of the pure energy eigenstates ${\cal S}_i$. The representations of ${\cal S}$, ${\cal S}_i$, ${\cal O}$, and $E$, are different for different gauges and models, but since by assumption the master equation has the fixed form given by Eq.~(\ref{me}), the correct dynamics are obtained from any given model provided that it correctly predicts the master equation rates, which are of the form $\gamma_{ijkl} = \kappa \bra{i}O\ket{j}\bra{k}O\ket{l}$. 
The exact gauge-invariant rates can be found using the non-truncated theory in any gauge $\alpha$ by letting $O=O_\alpha=R_{\alpha'\alpha}O_{\alpha'}R_{\alpha'\alpha}^\dagger$ and $H=H_\alpha=R_{\alpha'\alpha}H_{\alpha'}R_{\alpha'\alpha}^\dagger$ such that $\ket{i}=\ket{E_\alpha^i} = R_{\alpha'\alpha}\ket{E_{\alpha'}^i}$. Since $\ket{E_1^{2,i}}\approx \ket{E_1^i}$, the dipole gauge truncated model $H_1^2$ provides accurate approximations as $\gamma_{ijkl} \approx \kappa \bra{E_1^{2,i}}PO_1P\ket{E_1^{2,j}}\bra{E_1^{2,k}}PO_1P\ket{E_1^{2,l}}$. More generally, whether a given truncated model provides accurate predictions depends on the given model and on how it is used, as well as on the specific prediction sought and the mechanism of loss.

\begin{figure}[t]
\begin{minipage}{\columnwidth}
\begin{center}
\hspace*{-1mm}\includegraphics[width=\textwidth]{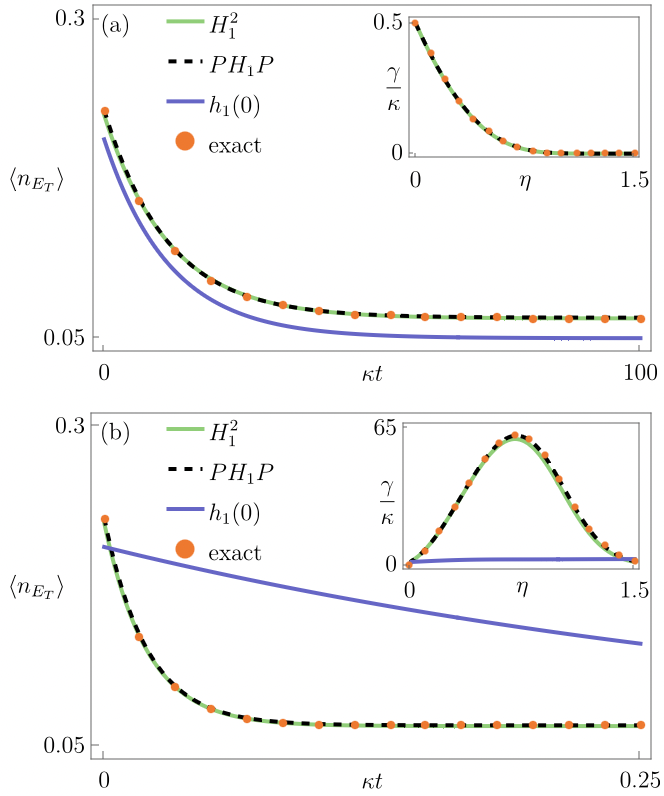}
\caption{The average photon numbers found using different two-level models and the exact theory, are plotted with time for the cases that dissipation occurs via (a) $Q_{E_{\rm T}}$, and (b) $p/\omega$. The initial state chosen is ${\cal S}_1$, and the coupling-strength chosen is $\eta=0.5$, while $\omega=\omega_0$ and $\mu\approx 70$. The insets show the corresponding decay rates as functions of the coupling strength $\eta$.}\label{Pi Dynamics}
\vspace*{-2mm}
\end{center}
\end{minipage}
\end{figure}

Suppose that as in Ref.~\cite{salmon_gauge-independent_2022}, loss occurs due to ``leakage" of the mode through a linear coupling to external modes of the transverse electric field quadrature $Q_{E_{\rm T}} = \sqrt{2v \over \omega}E_{\rm T}$. In the gauge $\alpha$ this observable is represented by the operator $iR_{0\alpha}(a^\dagger -a)R_{0\alpha}^\dagger$. Since the chosen observable $Q_{E_{\rm T}}$ is a linear function of $\Pi$ in the Coulomb gauge, the correct operator representation $i{\cal T}_{10}PR_{01}(a^\dagger -a)R_{01}^\dagger P{\cal T}_{10}^\dagger$ to use in conjunction with $h_1(0)$ does coincide with the Coulomb gauge representation $iP(a^\dagger-a)P$. Even here though, the solution $\rho$ of the resulting master equation is not a Coulomb gauge operator, because the average ${\rm tr}(\rho O'_0)$ of observable ${\cal O}'$ is inaccurate unless the Coulomb gauge representation $O'_0$ of ${\cal O}'$ also happens to satisfy $PO'_0P = {\cal T}_{10}PR_{01}O'_0 R_{01}^\dagger P{\cal T}_{10}^\dagger$. This is not the case if, for example, ${\cal O}'=n_{E_{\rm T}}$ whose average is plotted with time in Fig.~\ref{Pi Dynamics}(a). The $t\to \infty$ (stationary thermal state) average is shown for different temperatures as a function of coupling strength $\eta$ in Fig.~\ref{ph num}.

When considering a different loss mechanism by specifying a different observable ${\cal O}$ through which dissipation occurs, different behaviour may be observed. For example, choosing ${\cal O}$ such that $O_0=p/\omega$ provides a description of direct spontaneous emission into external modes resulting from a linear system-reservoir coupling in the Coulomb gauge. In this case the exact rates $\gamma_{ijkl} = \kappa \bra{E_0^i}p\ket{E_0^j}\bra{E_0^k}p\ket{E_0^l}$ are inaccurately predicted by $h_1(0)$ when treated as a Coulomb gauge model as shown in Fig.~\ref{Pi Dynamics}(b). This is because $p\not\approx {\cal T}_{10}PR_{01}p R_{01}^\dagger P{\cal T}_{10}^\dagger$. We conclude that in open system cases, as in closed system cases, the frame of $P{\cal H}$ that defines the model $h_1(0)$ cannot generally be treated as a truncated Coulomb gauge, otherwise incorrect predictions are obtained.

{\em Conclusions}. We have given a simple bound (\ref{bound3}) derived from the CSI, providing a general limit on the accuracy of light-matter theories that retain only a finite number of material energy levels. It is impossible in certain gauges to accurately approximate vectors and operators of interest using a theory restricted to only a few material energy levels. Trivially, once a gauge $g$ admitting accurate truncation is identified {\em every} subsequent unitary rotation ${\cal U}:P{\cal H}\to P{\cal H}$ defines an alternative frame in $P{\cal H}$ yielding identical predictions. Such constructions, although obviously unnecessary, have been pursued \cite{akbari_generalized_2023, dmytruk_gauge_2021, gustin_gauge-invariant_2023, gustin_what_2024,hughes_reconciling_2024, mercurio_flying_2022, mercurio_pure_2023, mercurio_regimes_2022, nodar_identifying_2023, salmon_gauge-independent_2022, savasta_gauge_2021, settineri_gauge_2021,garziano_gauge_2020,taylor_resolution_2020,mandal_theoretical_2023,taylor_resolving_2022,le_boite_theoretical_2020}, the apparent goal being to enable the direct use of arbitrary-gauge observable representations $O_{g'}$ after material truncation. The bound (\ref{bound3}) implies however that such use generally results in incorrect predictions. Thus, although it is of course possible to specify a gauge principle for the truncated theory $(P{\cal H},{\cal A}_P)$, frames within $P{\cal H}$ cannot be understood as gauges in the sense of the non-truncated theory. By way of example, we have studied the case of an anharmonic double-well dipole interacting resonantly with a single mode for which the standard dipole gauge QRM $H_1^2$ is known to yield certain predictions accurately. We have shown that a unitary rotation of this model, $h_1(0)$, proposed as a Coulomb gauge model in Ref.~\cite{di_stefano_resolution_2019} and treated as such in most subsequent literature, cannot be treated as a Coulomb gauge model lest it yields incorrect predictions.



\vspace*{-0.5cm}
\section*{SUPPLEMENTARY MATERIAL}

\subsection{Fidelities of dipole gauge truncated model eigenvectors}

\begin{figure}[t]
\begin{minipage}{\columnwidth}
\begin{center}
\hspace*{-1mm}\includegraphics[width=0.96\textwidth]{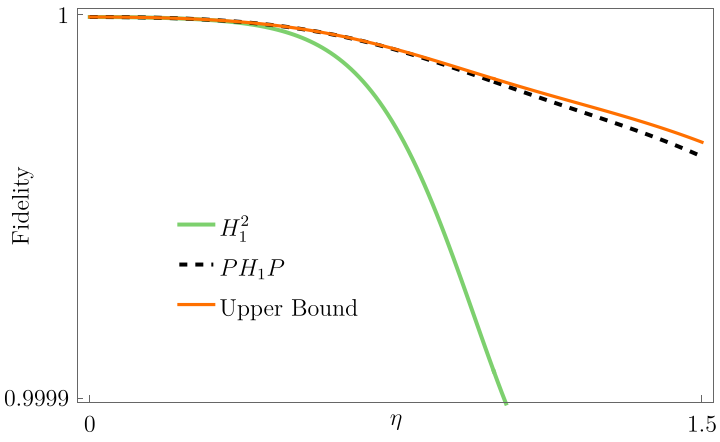}
\vspace*{-2mm}
\caption{The fidelities $F(E_1^{2,0},E_1^0)$ and $F({\tilde E}_1^{2,0},E_1^0)$ corresponding to the models $H_1^2$ and $PH_1P$, are shown as functions of coupling strength $\eta$. The upper bound is $\Vert PE_1^0\Vert^2 = F(PE_1^0/\Vert P E_1^0 \Vert,E_1^0)$.}\label{fid1}
\vspace*{5mm}
\hspace*{-1mm}\includegraphics[width=0.96\textwidth]{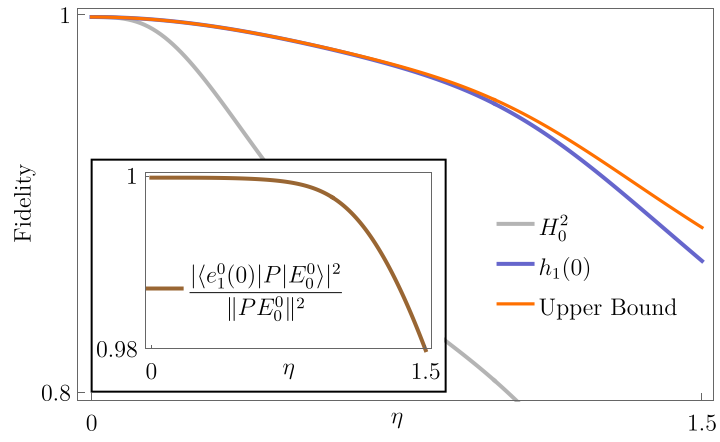}
\vspace*{-2mm}
\caption{The fidelities $F(E_0^{2,0},E_0^0)$ and $F(e_1^0(0),E_0^0)$ corresponding to the models $H_0^2$ and $h_1(0)$ are shown as functions of coupling strength $\eta$. The upper bound is $\Vert PE_0^0\Vert^2 = F(PE_0^0/\Vert P E_0^0 \Vert,E_0^0)$. The inset shows the ratio of the blue curve and the upper bound, which is the fidelity between $\ket{e_1^0(0)}$ and the optimal truncated state $P\ket{E_0^0}/\Vert PE_0^0 \Vert$. }\label{fid2}
\end{center}
\end{minipage}
\end{figure}

Here we show, via Fig.~\ref{fid1}, the fidelities $F(E_1^{2,0},E_1^0)$ and $F({\tilde E}_1^{2,0},E_1^0)$ as functions of coupling strength $\eta$, where $\ket{E_1^{2,0}}$ and $\ket{{\tilde E}_1^{2,0}}$ denote the ground eigenvectors of the dipole gauge truncated models $H_1^2$ and $PH_1P$ respectively. Both models yield fidelities close to unity over the coupling range shown, yet the fidelity of the model $H_1^2$ is seen to decay much more rapidly than that of $PH_1P$, which is in turn seen to increasingly deviate from the upper bound as $\eta$ increases.

Figure~\ref{fid2} shows the fidelities $F(E_0^{2,0},E_0^0)$ and $F(e_1^0(0),E_1^0)$ where $\ket{E_0^{2,0}}$ and $\ket{e_1^0(0)}$ are the ground eigenvectors of $H_0^2$ and $h_1(0)$ taken as Coulomb gauge two-level models. Since $F(E_1^{2,i},E_1^i)\approx 1$, and $h_1(0)={\cal T}_{10}H_1^2{\cal T}_{10}^\dagger$ where the generator of ${\cal T}_{10}$ is the projection onto $P{\cal H}$ of the generator of $R_{10}$, one might expect the normalised eigenvectors $\ket{e_1^i(0)}$ to approximate well the part of $\ket{E_0^i}=R_{10}\ket{E_1^i}$ that resides within $P{\cal H}$ once the latter is normalised, namely $P\ket{E_0^i}/\Vert PE_0^i\Vert$. Figure \ref{fid2} shows that this is the case for the ground state for moderate couplings. Noticeable deviations of  $F(e_1^0(0),PE_0^0/\Vert PE_0^0\Vert)$ from unity are observed for large $\eta$. Moreover, Figure \ref{fid2} shows that the upper bound $\Vert P E_0^0\Vert^2 = F(PE_0^0/\Vert PE_0^0\Vert,E_0^0)$ deviates significantly from unity. Thus, although $\ket{e_1^0(0)}$ can approximate $P\ket{E_0^0}/\Vert PE_0^0\Vert$ for moderate coupling strengths, the latter does not approximate $\ket{E_0^0}$ (because $Q\ket{E_0^0}$ is non-negligible, see Fig.~\ref{fid2} or Fig.~2 of the main text).

\subsection{Excited state population of the dipole defined relative to the Coulomb gauge}

\begin{figure}[H]
\begin{minipage}{\columnwidth}
\begin{center}
\vspace*{-5mm}
\hspace*{-1mm}\includegraphics[width=0.99\textwidth]{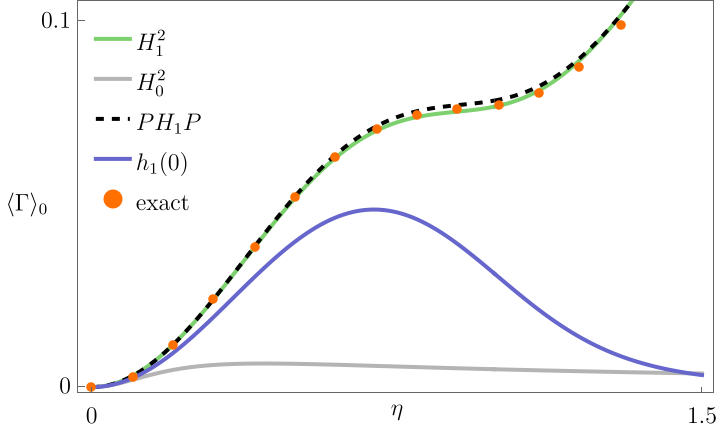}
\vspace*{-2mm}
\caption{The ground state population, $\langle \Gamma\rangle_0$, of the excited bare states of the dipole defined relative to the Coulomb gauge, is plotted with $\eta$ for various two-level models and the non-truncated theory. The model $h_1(0)$ is treated as a Coulomb gauge model. The parameters chosen are as in Fig. \ref{ph num}.}\label{matter excitation figure}
\end{center}
\end{minipage}
\end{figure}
Here we provide results for the excited state population of the dipole defined relative to the Coulomb gauge, that is, the total population $\Gamma$ of the excited states of the observable $K = {1\over 2m}(m{\dot x}+qA)^2+V(x)$, which is represented in the Coulomb gauge by the operator $H_m$. In the Coulomb gauge $\Gamma$ is represented by the operator $I-\ket{\epsilon_0}\bra{\epsilon_0}$. The models $H_1^2$ and $PH_1P$ yield accurate predictions whereas $h_1(0)$ taken as a Coulomb gauge model yields a generally incorrect average;
\begin{align}
\bra{e_1^0(0)}\left(I-\ket{\epsilon_0}\bra{\epsilon_0}\right)\ket{e_1^0(0)} \not\approx \langle \Gamma \rangle_0
\end{align}
Results for the ground state population are shown in Fig.~\ref{matter excitation figure}, and mirror those obtained for $n_{E_{\rm T}}$.

\subsection{Eigenvalue and transition spectra from different dipole gauge truncated models}

\begin{figure}[t]
\begin{minipage}{\columnwidth}
\begin{center}
\hspace*{-1mm}\includegraphics[width=0.96\textwidth]{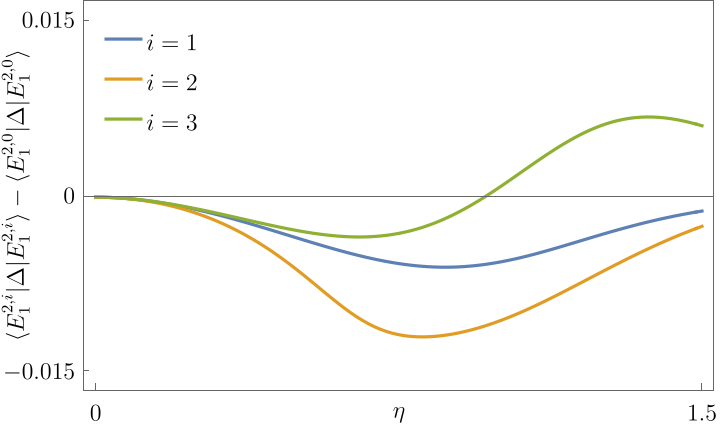}
\caption{The difference $\bra{E_1^{2,i}} \Delta \ket{E_1^{2,i}}-\bra{E_1^{2,0}} \Delta \ket{E_1^{2,0}}$ is plotted with coupling strength $\eta$ for $i=1,2,3$. The parameters chosen are as in Fig. \ref{ph num}. As expressed by inequality (\ref{approxsup}), each curve shows relatively small variations ($<0.02$) when compared with the variation in normalised energies shown in Fig.~\ref{Assume Correct Model 2}.}\label{ph num diff}
\vspace*{-2mm}
\end{center}
\end{minipage}
\end{figure}

Here we consider differences between averages found using different energy eigenvectors. The difference operator $\Delta$ given explicitly in Eq.~(\ref{diffnet}) of the main text can be expressed a number of ways. For example
\begin{align}
          \Delta &= \mathcal{T}_{01}a^\dagger a \mathcal{T}_{01}^\dagger - PR_{01}a^\dagger aR_{01}^\dagger P. \label{Expect Delta Q2}\\
          \Delta &={PH_1P-H_1^2\over \omega}\label{del2}
\end{align}
The first of these shows that $\bra{E_1^{2,i}} \Delta \ket{E_1^{2,i}}$ is the difference between the average of $n_{E_{\rm T}}$ in $i$'th energy state found using $h_1(0)$ treated as a Coulomb gauge model,
$\langle n_{E_{\rm T}} \rangle_i \approx \bra{e_1^i(0)}a^\dagger a \ket{e_1^i(0)}$, and the corresponding average found using $H_1^2$, which is $
\langle n_{E_{\rm T}} \rangle_i \approx \bra{E_1^{2,i}}PR_{01} a^\dagger a R_{01}^\dagger P\ket{E_1^{2,i}}$. The quantity $\bra{E_1^{2,i}} \Delta \ket{E_1^{2,i}}-\bra{E_1^{2,0}} \Delta \ket{E_1^{2,0}}$ is plotted as a function of coupling strength $\eta$ in Fig.~\ref{ph num diff} for the first three levels $i=1,2,3$. It is seen that $\bra{E_1^{2,i}} \Delta \ket{E_1^{2,i}}$ is relatively weakly dependent on $i$.

\begin{figure}[H]
\begin{minipage}{\columnwidth}
\begin{center}
\vspace*{2mm}
\hspace*{-1mm}
\includegraphics[width=1\textwidth]{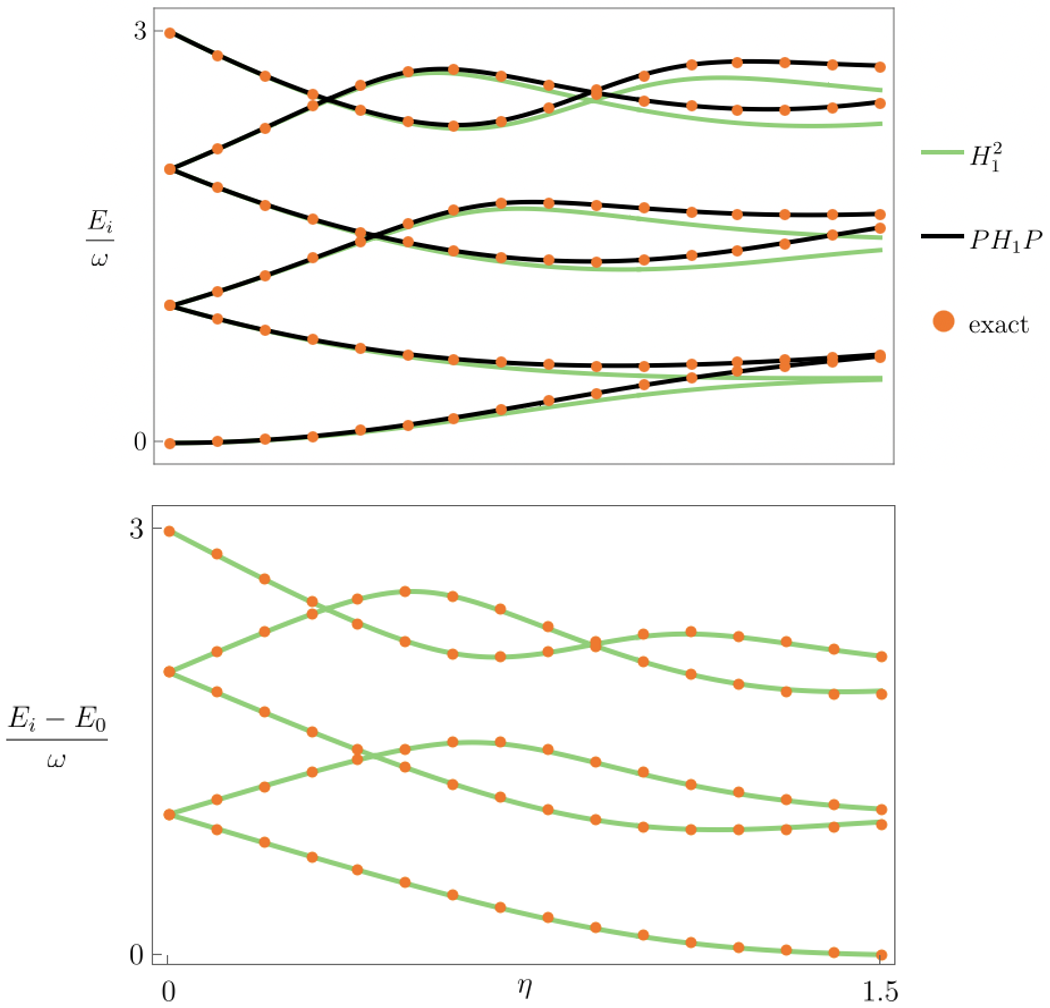}
\caption{The first six energies $E_i/\omega$ are plotted with $\eta$ (\textbf{Top}), with the predictions of $PH_1P$ found to be accurate unlike those of $H_1^2$. However, both models yield accurate normalised transition energies (\textbf{Bottom}). The parameters chosen are as in Fig. \ref{ph num}.}\label{Assume Correct Model 2}
\end{center}
\end{minipage}
\end{figure}

We now make use of this result together with the second expression for $\Delta$ given by Eq.~(\ref{del2}) to understand the behaviour of energies predicted by the different multipolar truncated models $PH_1P$ and $H_1^2$. The latter is representative of the entire class ${\cal C}_1$. Consider the average energy $E$ in the $i$'th energy eigenstate ${\cal S}_i$, which is nothing but the $i$'th energy eigenvalue; $\langle E \rangle_i=E_i$. The approximations of $E_i$ found using $PH_1P$ and $H_1^2$ are their respective $i$'th eigenvalues. A comparison, as functions of $\eta$, of these eigenvalues with $E_i$ is shown in Fig.~\ref{Assume Correct Model 2} for $i=0,...,5$. The model $PH_1P$ possesses eigenvalues that approximate the $E_i$ well. This could have been deduced from bound (\ref{bound3}) of the main text by noting that there exist vectors $\ket{\phi}\in P{\cal H}$ such that $F(\phi,E_1^i)\approx 1$. 

In particular, the eigenvectors of $PH_1P$ possess this property as do the eigenvectors $\ket{E_1^{2,i}}$ of $H_1^2$ (see Fig.~\ref{gpic} of the main text). However, the eigenvalues of $H_1^2$ do not in general approximate well the $E^i$, and nor therefore do those of any model within ${\cal C}_1$. In other words, although $F(E_1^{2,i},E_1^i)\approx 1$ and so $E_i = \bra{E_1^i}H_1\ket{E_1^i} \approx \bra{E_1^{2,i}}H_1\ket{E_1^{2,i}} = \bra{E_1^{2,i}}PH_1P\ket{E_1^{2,i}}$, this average is not in general well-approximated by $\bra{E_1^{2,i}}H_1^2\ket{E_1^{2,i}}=E_1^{2,i}$. The difference between these averages is $\omega\bra{E_1^{2,i}} \Delta \ket{E_1^{2,i}}$. 
However, as we have shown, $\bra{E_1^{2,i}} \Delta \ket{E_1^{2,i}}\approx \langle \Delta \rangle_i$ is only weakly dependent on $i$ (see Fig.~\ref{ph num diff}) and in particular
\begin{align}
\bra{E_1^{2,i}} \Delta \ket{E_1^{2,i}} - \bra{E_1^{2,0}} \Delta \ket{E_1^{2,0}} \ll {E_i -E_0\over \omega} \label{approxsup}
\end{align}
so that
\begin{align}
{E_i -E_0\over \omega} \approx  &{\bra{E_1^{2,i}}PH_1P\ket{E_1^{2,i}} \over \omega}- {\bra{E_1^{2,0}}PH_1P\ket{E_1^{2,0}}\over \omega} \nonumber \\ =& {E_1^{2,i}-E_1^{2,0}\over \omega}+ \bra{E_1^{2,i}} \Delta \ket{E_1^{2,i}} - \bra{E_1^{2,0}} \Delta \ket{E_1^{2,0}}\nonumber \\ \approx & {E_1^{2,i}-E_1^{2,0}\over \omega}
\end{align}
where the first equality follows from $F(E_1^{2,i},E_1^i)\approx 1$, the second from Eq.~(\ref{del2}), and the third from inequality~(\ref{approxsup}). We see therefore that $H_1^2$ does accurately predict normalised {\em transition} energies despite inaccurately predicting the $E_i$ themselves for large $\eta$ (Fig.~\ref{Assume Correct Model 2}).

\bibliographystyle{unsrt}
\bibliography{Bibliography}

\begin{thebibliography}{10}

\bibitem{PhysRev.49.324}
I.~I. Rabi.
\newblock On the process of space quantization.
\newblock {\em Phys. Rev.}, 49:324--328, Feb 1936.

\bibitem{JC_Model_Origional}
E.T. Jaynes and F.W. Cummings.
\newblock Comparison of quantum and semiclassical radiation theories with application to the beam maser.
\newblock {\em Proceedings of the IEEE}, 51(1):89--109, 1963.

\bibitem{de_bernardis_breakdown_2018}
Daniele De~Bernardis, Philipp Pilar, Tuomas Jaako, Simone De~Liberato, and Peter Rabl.
\newblock Breakdown of gauge invariance in ultrastrong-coupling cavity {QED}.
\newblock {\em Physical Review A}, 98(5):053819, 2018.

\bibitem{stokes_gauge_2019}
Adam Stokes and Ahsan Nazir.
\newblock Gauge ambiguities imply {Jaynes}-{Cummings} physics remains valid in ultrastrong coupling {QED}.
\newblock {\em Nature Communications}, 10(1):499, January 2019.

\bibitem{roth_optimal_2019}
Marco Roth, Fabian Hassler, and David~P. DiVincenzo.
\newblock Optimal gauge for the multimode {Rabi} model in circuit {QED}.
\newblock {\em Physical Review Research}, 1(3):033128, 2019.

\bibitem{di_stefano_resolution_2019}
Omar Di~Stefano, Alessio Settineri, Vincenzo Macrì, Luigi Garziano, Roberto Stassi, Salvatore Savasta, and Franco Nori.
\newblock Resolution of gauge ambiguities in ultrastrong-coupling cavity quantum electrodynamics.
\newblock {\em Nat. Phys.}, 15(8):803--808, August 2019.

\bibitem{gustin_gauge-invariant_2023}
Chris Gustin, Sebastian Franke, and Stephen Hughes.
\newblock Gauge-invariant theory of truncated quantum light-matter interactions in arbitrary media.
\newblock {\em Phys. Rev. A}, 107(1):013722, January 2023.

\bibitem{ashida_cavity_2021}
Yuto Ashida, Ata{\c c} {\.I}mamo{\u g}lu, and Eugene Demler.
\newblock Cavity {Quantum} {Electrodynamics} at {Arbitrary} {Light}-{Matter} {Coupling} {Strengths}.
\newblock {\em Physical Review Letters}, 126(15):153603, April 2021.
\newblock Publisher: American Physical Society.

\bibitem{stokes_implications_2022}
Adam Stokes and Ahsan Nazir.
\newblock Implications of gauge freedom for nonrelativistic quantum electrodynamics.
\newblock {\em Reviews of Modern Physics}, 94(4):045003, November 2022.
\newblock Publisher: American Physical Society.

\bibitem{taylor_resolution_2020}
Michael A.~D. Taylor, Arkajit Mandal, Wanghuai Zhou, and Pengfei Huo.
\newblock Resolution of {Gauge} {Ambiguities} in {Molecular} {Cavity} {Quantum} {Electrodynamics}.
\newblock {\em Phys. Rev. Lett.}, 125(12):123602, September 2020.

\bibitem{li_electromagnetic_2020}
Jiajun Li, Denis Golez, Giacomo Mazza, Andrew~J. Millis, Antoine Georges, and Martin Eckstein.
\newblock Electromagnetic coupling in tight-binding models for strongly correlated light and matter.
\newblock {\em Physical Review B}, 101(20):205140, May 2020.
\newblock Publisher: American Physical Society.

\bibitem{salmon_gauge-independent_2022}
Will Salmon, Chris Gustin, Alessio Settineri, Omar Di~Stefano, David Zueco, Salvatore Savasta, Franco Nori, and Stephen Hughes.
\newblock Gauge-independent emission spectra and quantum correlations in the ultrastrong coupling regime of open system cavity-{QED}.
\newblock {\em Nanophotonics}, 11(8):1573--1590, May 2022.

\bibitem{settineri_gauge_2021}
Alessio Settineri, Omar Di~Stefano, David Zueco, Stephen Hughes, Salvatore Savasta, and Franco Nori.
\newblock Gauge freedom, quantum measurements, and time-dependent interactions in cavity {QED}.
\newblock {\em Phys. Rev. Research}, 3(2):023079, April 2021.

\bibitem{arwas_metrics_2023}
Geva Arwas, Vladimir~E. Manucharyan, and Cristiano Ciuti.
\newblock Metrics and properties of optimal gauges in multimode cavity {QED}.
\newblock {\em Physical Review A}, 108(2):023714, August 2023.
\newblock Publisher: American Physical Society.

\bibitem{stokes_gauge-relativity_2023}
Adam Stokes, Hannah Riley, and Ahsan Nazir.
\newblock The {Gauge}-{Relativity} of {Quantum} {Light}, {Matter}, and {Information}.
\newblock {\em Open Systems \& Information Dynamics}, 30(03):2350016, September 2023.
\newblock Publisher: World Scientific Publishing Co.

\bibitem{savasta_gauge_2021}
Salvatore Savasta, Omar Di~Stefano, Alessio Settineri, David Zueco, Stephen Hughes, and Franco Nori.
\newblock Gauge principle and gauge invariance in two-level systems.
\newblock {\em Phys. Rev. A}, 103(5):053703, May 2021.

\bibitem{stokes_gauge_2020}
Adam Stokes and Ahsan Nazir.
\newblock Gauge non-invariance due to material truncation in ultrastrong-coupling {QED}.
\newblock {\em arXiv:2005.06499 [quant-ph]}, July 2020.

\bibitem{stokes_gauge_2024}
Adam Stokes and Ahsan Nazir.
\newblock Gauge non-invariance due to material truncation in ultrastrong-coupling quantum electrodynamics.
\newblock {\em Nature Physics}, 20(3):376--378, March 2024.
\newblock Publisher: Nature Publishing Group.

\bibitem{akbari_generalized_2023}
Kamran Akbari, Will Salmon, Franco Nori, and Stephen Hughes.
\newblock Generalized {Dicke} model and gauge-invariant master equations for two atoms in ultrastrongly-coupled cavity quantum electrodynamics.
\newblock {\em Phys. Rev. Research}, 5(3):033002, July 2023.

\bibitem{dmytruk_gauge_2021}
Olesia Dmytruk and Marco Schiró.
\newblock Gauge fixing for strongly correlated electrons coupled to quantum light.
\newblock {\em Phys. Rev. B}, 103(7):075131, February 2021.

\bibitem{gustin_what_2024}
Chris Gustin, Juanjuan Ren, and Stephen Hughes.
\newblock What is the spectral density of the reservoir for a lossy quantized cavity?, July 2024.
\newblock arXiv:2407.01855 [physics, physics:quant-ph].

\bibitem{hughes_reconciling_2024}
Stephen Hughes, Chris Gustin, and Franco Nori.
\newblock Reconciling quantum and classical spectral theories of ultrastrong coupling: role of cavity bath coupling and gauge corrections.
\newblock {\em Optica Quantum}, 2(3):133, June 2024.

\bibitem{mercurio_flying_2022}
Alberto Mercurio, Simone~De Liberato, Franco Nori, Salvatore Savasta, and Roberto Stassi.
\newblock Flying atom back-reaction and mechanically generated photons from vacuum, September 2022.
\newblock arXiv:2209.10419 [quant-ph].

\bibitem{mercurio_pure_2023}
Alberto Mercurio, Shilan Abo, Fabio Mauceri, Enrico Russo, Vincenzo Macrì, Adam Miranowicz, Salvatore Savasta, and Omar Di~Stefano.
\newblock Pure {Dephasing} of {Light}-{Matter} {Systems} in the {Ultrastrong} and {Deep}-{Strong} {Coupling} {Regimes}.
\newblock {\em Phys. Rev. Lett.}, 130(12):123601, March 2023.

\bibitem{mercurio_regimes_2022}
Alberto Mercurio, Vincenzo Macrì, Chris Gustin, Stephen Hughes, Salvatore Savasta, and Franco Nori.
\newblock Regimes of cavity {QED} under incoherent excitation: {From} weak to deep strong coupling.
\newblock {\em Phys. Rev. Research}, 4(2):023048, April 2022.

\bibitem{nodar_identifying_2023}
Álvaro Nodar, Ruben Esteban, Unai Muniain, Michael~J. Steel, Javier Aizpurua, and Mikołaj~K. Schmidt.
\newblock Identifying unbound strong bunching and the breakdown of the rotating wave approximation in the quantum {Rabi} model.
\newblock {\em Phys. Rev. Research}, 5(4):043213, December 2023.

\bibitem{garziano_gauge_2020}
Luigi Garziano, Alessio Settineri, Omar Di~Stefano, Salvatore Savasta, and Franco Nori.
\newblock Gauge invariance of the dicke and hopfield models.
\newblock {\em Phys. Rev. A}, 102:023718, Aug 2020.

\bibitem{mandal_theoretical_2023}
Arkajit Mandal, Michael~A.D. Taylor, Braden~M. Weight, Eric~R. Koessler, Xinyang Li, and Pengfei Huo.
\newblock Theoretical {Advances} in {Polariton} {Chemistry} and {Molecular} {Cavity} {Quantum} {Electrodynamics}.
\newblock {\em Chemical Reviews}, 123(16):9786--9879, August 2023.
\newblock Publisher: American Chemical Society.

\bibitem{taylor_resolving_2022}
Michael A.~D. Taylor, Arkajit Mandal, and Pengfei Huo.
\newblock Resolving ambiguities of the mode truncation in cavity quantum electrodynamics.
\newblock {\em Opt. Lett.}, 47(6):1446--1449, Mar 2022.

\bibitem{le_boite_theoretical_2020}
Alexandre Le~Boité.
\newblock Theoretical methods for ultrastrong light–matter interactions.
\newblock {\em Advanced Quantum Technologies}, 3(7):1900140, 2020.

\end{thebibliography}

\end{document}